%% file: rot_cy-8b.tex
\documentclass[fleqn,11pt]{article}
\usepackage{amsmath,amssymb,amsfonts,latexsym}
\usepackage{graphicx}

\usepackage{color}

\def\eqn#1{\eq\eqref{#1}}
\def\rf{\eqref}
\def\mn{_{\mu\nu}}
\def\MN{^{\mu\nu}}
\def\mN{_\mu^\nu}

\def\R{{\mathbb R}}
\def\tK{{\widetilde K}}
\def\tT{{\widetilde T}}

\def\kappa{\varkappa}

\def\grav{gravitational}
\def\GR{general relativity}
\def\cy{cylindrical}
\def\cyl{cylindrically symmetric}

\def\wh{wormhole}
\def\whs{wormholes}
\def\asflat{asymptotically flat}

\def\elmag{electromagnetic}

\input preamble.tex

\begin{document}
\thispagestyle{empty}
\twocolumn[

\Title{Rotating Melvin-like universes and wormholes in general relativity}

\Aunames{K. A. Bronnikov,\au{a,b,c,1} V. G. Krechet,\au{d} and V. B. Oshurko\au{d}}
		
\Addresses{
\adr a {\small Center for Gravitation and Fundamental Metrology, VNIIMS,
             Ozyornaya ul. 46, Moscow 119361, Russia}
\adr b {\small Peoples' Friendship University of Russia (RUDN University), 
             ul. Miklukho-Maklaya 6, Moscow 117198, Russia}
\adr c {\small National Research Nuclear University ``MEPhI''
                    (Moscow Engineering Physics Institute), Moscow, Russia}
\adr d {\small Moscow State Technological University ``Stankin'',
         Vadkovsky per. 3A, Moscow 127055, Russia}  
	}		
	
\Abstract
  {We find a family of exact solutions to the Einstein-Maxwell equations for rotating cylindrically
   symmetric distributions of a perfect fluid with the equation of state $p = w\rho$ ($|w| < 1$),  carrying 
   a circular electric current in the angular direction. This current creates a magnetic field along the $z$ 
   axis. Some of the solutions describe geometries resembling that of Melvin's static  magnetic universe 
   and contain a regular symmetry axis, while some others (in the case $w > 0$) describe traversable 
   wormhole geometries which do not contain a symmetry axis. Unlike Melvin's solution, those with
   rotation and a magnetic filed cannot be vacuum and require a current. The wormhole solutions 
   admit matching with flat-space regions on both sides of the throat, thus forming a cylindrical wormhole
   configuration potentially visible for distant observers residing in flat or weakly curved parts of space. 
   The thin shells, located at junctions between the inner (wormhole) and outer (flat) regions, consist 
   of matter satisfying the Weak Energy Condition under a proper choice of the free parameters of the 
   model, which thus forms new examples of phantom-free wormhole models in general relativity. 
   In the limit $w \to 1$, the magnetic field tends to zero, and the wormhole model tends to the one   
   obtained previously, where the source of gravity is stiff matter with the equation of state $p = \rho$.
  } 
	
] 
\email 1 {kb20@yandex.ru}

\section{Introduction}

  Cylindrical symmetry is the second (after the spherical one) simplest space-time symmetry making it
  possible to obtain numerous exact solutions in \GR\ and its extensions, characterizing local strong 
  gravitational field configurations. One of the motivations of studying \cyl\ configurations is the possible 
  existence of such linearly extended structures as cosmic strings as well as the observed cosmic jets.
  A large number of static \cyl\ solutions have been obtained and studied since the advent of \GR, 
  including vacuum,  electrovacuum, perfect fluid and others, see reviews in \cite{exact-book, BSW19}
  and references therein.  

  Melvin's famous solution to the Einstein-Maxwell equations, an ``electric or magnetic geon'' \cite{melvin},
  is a completely regular static, \cyl\ solution with a longitudinal electric or magnetic field as the only source 
  of gravity. It is a special case from a large set of static \cyl\ Einstein-Maxwell fields, see more details in 
  \cite{kb79, BSW19}.   

  An important distinguishing feature of \cy\ symmetry as compared to the spherical one is the possible inclusion  
  of rotation, avoiding complications inherent to the more realistic axial symmetry, not to mention the general 
  nonsymmetric space-times.  Accordingly, a great number of exact stationary (assuming rotation) 
  solutions to the Einstein equations are known, with various sources of gravity: the cosmological constant
  \cite{lanczos, lewis, vac-Lam1, vac-Lam2, vac-Lam3}; scalar fields with different self-interaction potentials 
  \cite{BLem13,BK15, erices}; rigidly or differentially rotating dust \cite{van_stockum, bonnor09, iva02c}, 
  dust with electric charge \cite{iva02b} or a scalar field \cite{santos82}, fluids with different equations of 
  state, above all, perfect fluids with $p=w\rho$, $w=\const$ (in usual notations)
  \cite{hoen79, davids97, davids00, skla99, iva02a}, some kinds of anisotropic fluids 
  \cite{anis1, anis2, anis3, anis4} etc., see also references therein and the reviews  \cite{exact-book, BSW19}. 
  
  In this paper we obtain rotating counterparts of the static \cyl\ solutions to the Einstein-Maxwell 
  equations with a longitudinal magnetic field. It turns out that such a field cannot exist without a source 
  in the form of an electric current, and we find solutions where such a source is a perfect fluid with 
  $p = w\rho$. Many features of these solutions are quite different from those of the static ones, 
  in particular, their common feature is the emergence of closed timelike curves at large radii. 
  Also, there is a family of \wh\ solution that do not have a symmetry axis but contain a throat as
  a minimum of the circular radius. As in our previous studies \cite{BLem13, BK15, BK18, BBS19}, 
  we try to make such \whs\ potentially observable from spatial infinity by joining outer flat-space regions   
  at some junction surfaces and verify the validity of the Weak Energy Condition for matter residing on 
  these surfaces. 
  
  The structure of the paper is as follows. Section 2 briefly describes the general formalism, in Section 3 
  we find solutions of the field equations, in Section 4 we discuss the properties of Melvin-like solutions,
  and in Section 5 the \wh\ family. Section 6 contains some concluding remarks.  
  
\section{Basic relations}

  We consider stationary \cyl\ space-times with the metric
\bearr                                                    \label{ds-rot}
         ds^2 = \e^{2\gamma(x)}[ dt - E(x)\e^{-2\gamma(x)}\, d\varphi ]^2- \e^{2\alpha(x)}dx^2 
\nnnv \inch
	- \e^{2\mu(x)}dz^2 - \e^{2\beta(x)}d\varphi^2,
\ear
  where $x^0 = t\in \R$, $x^1 =x$, $x^2 = z\in \R$ and $x^3 =\varphi\in [0, 2\pi)$ are the
  temporal,  radial, longitudinal and angular (azimuthal) coordinates, respectively. The variable 
  $x$ is here specified up to a substitution $x \to f(x)$, therefore its range depends on both
  the geometry itself and the ``gauge''  (the coordinate condition). The off-diagonal 
  component $E$  describes rotation, or the vortex component of the gravitational field.
  In the general case, this vortex \grav\ field is determined by the 4-curl
  of the orthonormal tetrad field  $e^\mu_m$ (Greek and Latin letters are here assigned to
  world and tetrad indices, respectively) \cite{kr2, kr4}:
\beq                        \label{om1}
	\omega^\mu = \Half \eps^{\mu\nu\rho\sigma} e_{m \mu} \d_\rho e^m_\sigma. 	   	
\eeq   
  Kinematically, the axial vector $\omega^\mu$ is the angular velocity of tetrad rotation,
  it determines the proper angular momentum density of the gravitational field,
\beq
		S^\mu(g) = \omega^\mu/\kappa, \qquad \kappa = 8\pi G,
\eeq  
  where $G$ is the Newtonian \grav\ constant. In space-times with the metric \rf{ds-rot} 
  we have 
\beq                        \label{om2}
            \omega^\mu = \Half \delta^{\mu 2} (E\e^{-2\gamma})' \e^{\gamma-\alpha-\beta-\mu} 
\eeq    
  (a prime stands for $d/dx$), and it appears sufficient to consider its absolute value 
  $\omega(x) = \sqrt{\omega^\mu \omega_\mu}$ that has the meaning
  of the angular velocity of a congruence of timelike curves (vorticity) \cite{BLem13, kr2, kr4},
\beq                                                  \label{om}
          \omega = \Half (E\e^{-2\gamma})' \e^{\gamma-\beta-\alpha}.
\eeq
  Furthermore, in the reference frame comoving to matter as it rotates in the 
  azimuthal ($\varphi$) direction,  the stress-energy tensor (SET) component
  $T^3_0$ is zero, therefore due to the Einstein equations the Ricci tensor 
  component $R_0^3 \sim (\omega \e^{2\gamma+\mu})' = 0$, which leads to \cite{BLem13}
\beq       	      					\label{omega}
	\omega = \omega_0 \e^{-\mu-2\gamma}, \cm \omega_0 = \const.
\eeq
  Then, according to \rf{om}, 
\beq                          \label{E}
	E(x) = 2\omega_0 \e^{2\gamma(x)} \int \e^{\alpha+\beta-\mu-3\gamma}dx.
\eeq
  Note that \eqs \rf{om2}--\rf{E} are valid for an arbitrary choice of the radial coordinate $x$. 
  Preserving this arbitrariness, we can write the nonzero components of the Ricci $(R\mN)$ 
  tensor as 
 \bear                 \label{Ric}
      R^0_0 \eql -\e^{-2\alpha}[\gamma'' + \gamma'(\sigma' -\alpha')] - 2\omega^2,
\nnv      
      R^1_1 \eql -\e^{-2\alpha}[\sigma'' + \sigma'{}^2 - 2U - \alpha'\sigma']+ 2\omega^2,
\nnv      
      R^2_2 \eql -\e^{-2\alpha}[\mu'' + \mu'(\sigma' -\alpha')],  
\nnv      
      R^3_3 \eql -\e^{-2\alpha}[\beta'' + \beta'(\sigma' -\alpha')] + 2\omega^2, 
\nnv
      R^0_3 \eql G^0_3 =  E \e^{-2\gamma}(R^3_3 - R^0_0), 
\ear
 where we are using the notations
\beq
            \sigma = \beta + \gamma + \mu, \qquad
            U = \beta'\gamma'  + \beta'\mu' + \gamma' \mu'.
\eeq     
  The Einstein equations may be written in two equivalent forms 
\bearr    \label{EE1}
            G\mN \equiv R\mN - \half \delta\mN R = -\kappa T\mN, \quad {\rm or}
\\   \lal                 \label{EE2}
            R\mN = - \kappa \tT\mN \equiv -\kappa(T\mN - \half \delta\mN T).          
\ear    
  $R$ being the Ricci scalar and $T$ the SET trace. We will mostly use the form \rf{EE2},
  of the equations, but it is also necessary to write the constraint equation from \rf{EE1}, 
  which contains only first-order derivatives of the metric and represents a first integral 
  of the other equations:
\beq                  \label{G11}
	      G^1_1 = \e^{-2\alpha} U + \omega^2 = - \kappa T^1_1.
\eeq    
   Owing to the last line of \rf{Ric} and its analogue for $T\mN$, in the Einstein equations 
   it is sufficient to solve the diagonal components, and then their only off-diagonal component 
   holds automatically \cite{BLem13}.  
    
  As is evident from \rf{Ric}, the diagonal components of both the Ricci ($R\mN$) and the 
  Einstein ($G\mN$) tensors split into the corresponding tensors for the static metric 
  (the metric (\ref{ds-rot}) with $E=0$) plus a contribution containing $\omega$ \cite{BLem13}: 
\bearr 		\label{R-omega}
    		R\mN = {}_s R\mN + {}_\omega R\mN, 
\nnn                     \cm
	{}_\omega R\mN = \omega^2 \diag (-2, 2, 0, 2),  
\yyy     		\label{G-omega}  
		G\mN = {}_s G\mN + {}_\omega G\mN,  
\nnn                      \cm
   	{}_\omega G\mN = \omega^2 \diag (-3, 1, -1, 1),  
\ear
  where ${}_s R\mN$ and ${}_s G\mN$ correspond to the static metric. It turns out that
  the tensors ${}_s G\mN$ and ${}_\omega G\mN$ (each separately) obey the conservation 
  law $\nabla_\alpha G^\alpha_\mu =0$ in terms of this static metric. Therefore, the tensor 
  ${}_\omega G\mN/\kappa$ may be interpreted as the SET of the vortex \grav\ field.
  It possesses quite exotic properties (thus, the effective energy density is 
  $ -3\omega^2/\kappa <  0$), which favor the existence of wormholes, and indeed,
  a number of \wh\ solutions with the metric \rf{ds-rot} have already been obtained
  \cite{BLem13,BK15,kr4,BBS19} with sources in the form of scalar fields, isotropic or
  anisotropic fluids. Further on we will obtain one more solution of this kind, now 
  supported by a perfect fluid and a magnetic field due to an electric current.
  
\section{Solutions with a perfect fluid and a magnetic field}
\subsection{The \elmag\ field. A no-go theorem}

  Consider a longitudinal ($z$-directed) magnetic field, corresponding to the 4-potential
\beq                        \label{A_phi}
		A_\mu = (0, 0, 0, \Phi(x)),
\eeq      
  so that the only nonzero components of the Maxwell tensor $F\mn = \d_\mu A_\nu - \d_\nu A_\mu$ 
  are $F_{13} = -F_{31} = \Phi'(x)$. The nonzero contravariant components $F\MN$ are
\beq     \nhq                  \label{Fmn}
		F^{13} = \e^{-2\alpha-2\beta}\Phi', \quad F^{01}= E\e^{-2\alpha-2\beta-2\gamma}\Phi'.
\eeq    
  The magnetic field magnitude (magnetic induction) $B$ is determined by $B^2 = F^{13}F_{13}$,
  so that the \elmag\ field invariant is $F\mn F\MN = 2 B^2$.
  
\medskip\noi  
  {\bf No-go theorem.}
  It can be shown that a free longitudinal magnetic field is incompatible with a nonstatic ($\omega \ne 0$)
  metric \rf{ds-rot}. This follows from solving the Maxwell equations, which, for $F\MN$ of the 
  form \rf{Fmn} read
\beq                          \label{Max0}
		(\sqrt{-g} F^{13})' =0, \qquad   (\sqrt{-g} F^{01})' =0,  
\eeq    
  where 
\[  
		g = \det (g\mn), \qquad		\sqrt{-g}= \e^{\alpha+\beta+\gamma+\mu}. 
\]  
   Equations \rf{Max0} are integrated to give, respectively,
\beq         \nhq   					\label{Phi'}
		   \Phi' \e^{-\alpha-\beta+\gamma+\mu} = h_1,\quad
		   E \Phi' \e^{-\alpha-\beta-\gamma+\mu} = h_2,
\eeq      
   where $h_1, h_2 = \const \ne 0$. From \rf{Phi'} it follows $E \e^{-2\gamma} = h_2/h_1$, 
   whence  $(E \e^{-2\gamma})' =0$, and according to \rf{omega}, $\omega =0$. We have shown that 
   a free longitudinal magnetic field cannot support a vortex \grav\ field with the metric \rf{ds-rot}. 
   
   (Let us also note that in the case $E \e^{-2\gamma} =E_1=\const$, the term 
   $E \e^{-2\gamma}d\varphi$ is eliminated from \rf{ds-rot} by introducing the new time coordinate
   $t' = t - E_1\varphi$, making the metric explicitly static.)

\subsection{The fluid}

  To circumvent the above no-go theorem, that is, to avoid the relation $E \e^{-2\gamma} = \const$,
  let us introduce a source of the magnetic field in the form of an electric current density 
  $J^\mu = \rho_e u^\mu$, where $\rho_e$ is the effective electric charge density,\footnote
  	{If we introduce a real nonzero charge density, it becomes necessary to consider, in addition,
  	a Coulomb electric field, which will make the problem hardly tractable. We therefore consider
  	an azimuthal electric current as if in a coil, in a neutral medium like a conductor with free 
  	electrons and ions at rest.  }
  and $u^\mu$  is the 4-velocity satisfying the usual normalization condition $u_\mu u^\mu =1$. 
  We will assume that the effective charge distribution is at rest in our rotating reference frame, 
  so that 
\beq                                       \label{J}
		u^\mu = (\e^{-\gamma}, 0,0,0), \quad J^\mu =  (\rho_e \e^{-\gamma}, 0,0,0).
\eeq    
  As usual, the electric charge conservation equation $\nabla_\mu J^\mu$ holds automatically
  due to the Maxwell equations $\nabla_\nu F\MN = J^\mu$.
  
  Two nontrivial Maxwell equations now read
\bearr                          \label{Max1}
		(\sqrt{-g} F^{13})' =0,  
\yyy                          \label{Max2}
		\frac{1}{\sqrt{-g}} (\sqrt{-g} F^{01})' = \rho_e \e^{-\gamma}.
\ear
  Integrating \eqn{Max1}, we obtain, as before,
\beq
		\Phi' = h \e^{\alpha+\beta-\gamma-\mu}, \qquad h = \const,
\eeq
  and substituting this $\Phi'$ to \eqn{Max2} with \rf{Fmn}, we arrive at the following expression 
  for $\rho_e$:
\beq                       \label{rho0}
		\rho_e = h \omega_0 \e^{-2\mu - 3\gamma}.
\eeq    

  On the other hand, the electric charges should have a material carrier, for which we will assume
  a perfect fluid with a barotropic equation of state and postulate a constant ratio of the effective
  charge density $\rho_e$ to energy density $\rho$:
\beq                      \label{flu}
		\rho_e /\rho= A = \const; \qquad  p/\rho = w=\const,
\eeq    
  $p$ being the fluid pressure. We do not fix the value of $w$ but later on we will obtain a restriction 
  on it. The fluid must obey the conservation law $\nabla_\nu T\mN =0$, which gives in our 
  comoving reference frame
\beq                        \label{cons}
		p'+ (p+\rho) \gamma' = 0,  
\eeq    
  which, for $w\ne 0$, leads to the expression 
\beq                         \label{rho1}
		\rho = \rho_0 \e^{-\gamma(w+1)/w}, \quad\ \rho_0 = \const.
\eeq            
   Comparing  \rf{rho0} and \rf{rho1}, taking into account the assumption $\rho_e /\rho= A = \const$,
   we obtain a relation between the metric coefficients $\e^{2\gamma}$ and $\e^{2\mu}$:
\beq                         \label{ga-mu}
	        A \rho_0 \e^{2\mu} = h\omega_0 \e^{- \gamma(w+1)/(2w)}.	        		
\eeq
   In the case $w=0$ (zero pressure), the conservation law \rf{cons} simply leads to 
   $\gamma = \const$.   
   
\subsection{Solution of the Einstein equations}

  To address the Einstein equations, let us write the expressions for the SETs of the perfect fluid
  and the electromagnetic field. For the fluid we have 
\beq
		T\mN[f] = \rho \diag(1, -w, -w, -w).
\eeq    
  For the \elmag\ field SET we have the standard expression 
\[
		T\mN[e] = \frac{1}{16\pi}\big(- F_{\mu\alpha}F^{\nu\alpha} 
						+ 4\delta\mN F_{\alpha\beta}F^{\alpha\beta} \big),
\]
  which in our case leads to
\beq
		T\mN[e] = \frac{B^2}{8\pi} \diag(1. -1, 1, -1) \oplus T^0_3[e],
\eeq    
  where $B^2 =  h^2 \e^{-2\gamma-2\mu}$, and the only off-diagonal component 
\beq
		T^0_3[e] = - \frac 1{4\pi} h \Phi' \e^{-2\gamma}
\eeq    
  does not affect the solution process, as mentioned in a remark after \eqn{G11}.      
  
  Thus far all relations and expressions were written in terms of an arbitrary radial 
  coordinate $x$. However, to solve the Einstein equations it is helpful, at last, to
  choose the ``gauge'', and by analogy with our previous studies we will use the 
  harmonic radial coordinate corresponding to
\beq                    \label{harm}
		\alpha = \beta + \gamma + \mu.
\eeq		
    
  With the above expressions for the SET and \eqs \rf{Ric}, the noncoinciding components of
  \eqs \rf {EE2} and \rf{G11}, taking into account the expressions \rf{omega} for $\omega$
  and  \rf{rho1} for $\rho$, may be written as 
\bearr             \label{00}
	\gamma'' = - 2\omega_0^2 \e^{2\beta -2\gamma}
				 + \frac{1+3w}2 K\e^{2\beta-\gamma}	
\nnn \inch				 
				 + Gh^2 \e^{2\beta},  
\\ \lal                \label{22}
         \mu'' =  \frac{w - 1}2 K\e^{2\beta-\gamma} + Gh^2 \e^{2\beta},  
\\ \lal                 \label{33}
         \beta'' =   2\omega_0^2 \e^{2\beta -2\gamma} 
         		+  \frac{w - 1}2 K\e^{2\beta-\gamma} - Gh^2 \e^{2\beta},  
\\ \lal                  \label{11}
         \beta'\gamma' + \beta' \mu' + \gamma'\mu' = -\omega_0^2 \e^{2\beta -2\gamma} 
\nnn \inch         
		         		+  w K\e^{2\beta-\gamma} + Gh^2 \e^{2\beta}  ,
\ear    
  where $K = \kappa h \omega_0/A$  (recall that $\kappa = 8\pi G$).
  
  Now, combining \eqs \rf{00} and \rf{22} with \rf{ga-mu}, we arrive at the algebraic 
  equation for $\gamma$ with constant coefficients,
\bearr                   \label{C1}
		4\omega_0^2 (1-2w)\e^{-2\gamma} +(8w^2 -3w -1)K\e^{-\gamma}
\nnn \inch		
					+ 2(4w-1) Gh^2 =0, 
\ear    
  from which it follows $\gamma = \const$, and we can without loss of generality put 
  $\gamma \equiv 0$ by choosing a time scale. Then \eqn{ga-mu} implies $\mu = \const$
  which also allows us to put $\mu \equiv 0$ by choosing the scale along $z$; 
  from \rf{ga-mu} we then obtain a relation among the constants: $h\omega_0 =A\rho_0$,
  so that, in particular, $K = \kappa \rho_0$.
  
  With $\e^\gamma = \e^\mu =1$, from \eqs \rf{00} and \rf{22} we find
\beq                     \label{C2}
			Gh^2 = \Half (1-w)K, \qquad \omega_0^2 = \Half(1+w)K, 
\eeq    
  which leads to a conclusion on the range of $w$:
\beq
                           -1 < w < 1.
\eeq    
  With \rf{C2} it is directly verified that \eqs \rf{11} and \rf{C1} also hold. All our constant
  parameters may be expressed in terms of two of them, for example, $\omega_0$ and $w$:
\bearr
		Gh^2 = \frac{1-w}{1+w} \omega_0^2, \qquad
		\kappa\rho_0 = \frac{2\omega_0^2}{1+w}, 
\nnn \cm
		A = \frac{\rho_e}{\rho} = 4\pi\sqrt{G(1-w^2)}.  
\ear      
     
   We see that in our system not only $\mu =\gamma =0$, but also both densities 
   $\rho$ and $\rho_e$ as well as the angular velocity $\omega$ are constant. 
   It is also of interest that the two limiting cases of the equation of state, $w=1$
   (maximum stiffness compatible with causality) and $w=-1$ (the cosmological constant)
   are excluded in the present system. In both these cases, static and stationary \cyl\
   solutions without an \elmag\ field are well known 
   \cite{lewis,BSW19, vac-Lam1,BLem13,erices,BBS19,anis4}.
   
   The remaining differential equation 	\rf{33} has the Liouville form
\beq                     \label{liu}
	   \beta'' = \frac{4w}{w+1} \omega_0^2 \e^{2\beta}.	
\eeq      
   and has the first integral
\beq
	\beta'{}^2 = \frac{4w}{w+1} \omega_0^2 \e^{2\beta} + k^2 \sign k,
\eeq		
  with $k = \const$. The further integration depends on the signs of $w$ and $k$:
\bearr  \nhq          \label{be1}
	{\bf 1.} \ \ w < 0,\  k>0, \quad     \e^{\beta} =  \frac k {m \cosh (kx)};.
\yyy \nhq          \label{be2}
	{\bf 2.} \ \ w>0, \ k>0:   \quad	    \e^{\beta} =  \frac k {m \sinh (kx)};.  
\yyy \nhq          \label{be3}
	{\bf 3.} \ \ w>0, \ k=0:   \quad	    \e^{\beta} =  \frac 1{m x};.
\yyy \nhq          \label{be4}
	{\bf 4.} \ \ w>0, \ k<0:   \quad	    \e^{\beta} =  \frac {|k|} {m \cos (|k|x)};.
\ear	
   where we have denoted  $m = \bigg(\dfrac{4 |w| \omega_0^2}{w+1}\bigg)^{1/2}$.

   In the previously excluded case $w =0$ (dustlike matter), the equality $\gamma =\const$ is 
   immediately obtained from \rf{cons}, $\mu  = \const$ then follows from \rf{ga-mu},
   and as before, without loss of generality, we can put $\mu = \gamma =0$. 
   Instead of \rf{liu}, we obtain $\beta'' =0$ whence we can write
\beq               \label{be0} 
   	\e^\beta = r_0 \e^{kx}, \qquad  r_0, k =\const.
\eeq   
   In all cases the off-diagonal metric function $E$ is easily obtained as
\beq                       \label{E1}
		E(x) = 2\omega_0 \int \e^{2\beta} dx.
\eeq      

\section{Melvin-like universes}

 Melvin's electric or magnetic geon \cite{melvin} is among the most well-known static, \cyl\ solutions to 
  the Einstein-Maxwell equations; it is a special solution from a large class of static, \cyl\ solutions
  with radial, azimuthal and longitudinal electric and/or magnetic fields, see, e.g., 
  \cite{kb79, exact-book, BSW19}. Its metric may be written in the form \cite{kb79,BSW19}
\bearr      \label {ds-melvin}
			ds^2 = (1+ q^2x^2)^2 (d t^2 - dx^2 - d z^2 )
\nnn \inch			
			- \frac {x^2}{(1+ q^2 x^2)^2} d\varphi^2, 
\ear
  where $x \geq 0$, and the magnetic (let us take it for certainty) field magnitude is 
\[
  		B = B_z = 2q(1+q^2 x^2)^{-2},
\]  
  with $q = \const$ characterizing the effective current that might be its source. 
  However, this solution describes a purely field configuration existing without any massive 
  matter, electric charges or currents. Both the metric and the magnetic field are regular on the axis $x=0$. 
  The other ``end'', $x \to \infty$, is infinitely far away (the distance $\int \sqrt{-g_{xx}}dx$ diverges), the 
  magnetic field vanishes there, and the circular radius $ r = \sqrt{-g_{\varphi\varphi}}$ also tends to zero, 
  so that the whole configuration is closed in nature, without spatial infinity, and with finite total magnetic 
  field energy per unit length along the $z$ axis.   

  As we saw in Sec. 3.1, such a free magnetic field cannot support a 
  rotating counterpart of Melvin's solution, but Einstein-Maxwell solutions with a longitudinal 
  magnetic field are obtained in the presence of perfect fluids with electric currents. Let us briefly 
  discuss their main features.    
  
  In all cases under consideration, the magnetic field is directed along the $z$ axis and has the 
  constant magnitude $B = h$, while the metric has the form 
\beq                                     \label{ds-m}
	  ds^2 = (dt - E d\varphi)^2 - \e^{2\beta} dx^2  - dz^2 - \e^{2\beta} d\varphi^2, 
\eeq    
  and $E$ is determined by \eqn{E1}. Note that both $\varphi$ and $x$ are dimensionless while
  $t, z$ and $\e^\beta$ have the dimension of length. 
  
\subsection*{Dustlike matter, \eqn{be0}}  
 \def\oo{\omega_0}
 \def\ok{{\bar k}}
 
  Let us begin with the case $w=0$. For $E(x)$ we find
\bearr                          \label{E0}
       E = E_0 + \frac {\oo r_0^2}{k} \e^{2kx} = E_0 + \frac{\oo}{k} r^2, 
\nnn 
	r = r_0\e^{kx},   \qquad   E_0 = \const,	        
\ear  
 where $E_0$ is an integration constant. In terms of the coordinate $r$, the metric reads
\beq                                     \label{ds0}
	ds^2 = \Big(dt - E d\varphi^2\Big)^2 - k^{-2} dr^2 -dz^2 - r^2 d\varphi^2.	 
\eeq 
  The symmetry axis $r=0$ is regular in the case $E_0=0$, $k=1$.\footnote
  		{The axis regularity conditions require \cite{exact-book, BSW19,ws-book} finite values of the 
  		curvature invariants plus local flatness (sometimes also called ``elementary flatness'') as a correct 
           	circumference to radius ratio for small circles around the axis, which in our case 
  		leads to the condition $\e^{-2\alpha} R'{}^2 \to 1$, where $R = \sqrt{-g_{33}}$.}
   Also, in this case
\beq    \label{g33-0}
  	g_{33} = E^2 - r^2 = r^2 ( -1 + \oo^2 r^2)	
\eeq     
  changes its sign at $r = \omega^{-1}$, and at larger $r$ the lines of constant $t,r,z$
  (coordinate circles) are timelike, thus being closed timelike curves (CTCs) violating 
  causality.
   
\subsection*{Solution 1, \eqn{be1}}  

  For $w < 0$,  with \rf{be1}, for $E(x)$ we calculate
\beq                      \label{E1}
	E = E_0 + \frac{2 k \oo}{m^2} \tanh kx,		
\eeq     
  The metric has the form 
\bearr                                     \label{ds1}
	  ds^2 = (dt - E d\varphi)^2 - dz^2 
\nnn \cm	  
	  -  \frac {k^2} {m^2 \cosh^2 (kx)} (dx^2 + d\varphi^2), 	
\ear  
 where, for convenience, we have rearranged the terms with $dz^2$ and $dx^2$ as compared to \rf{ds-m}.

  For $g_{33}$, similarly to \rf{g33-0}, again putting $E_0=0$ and recalling the definition of $m$, we obtain
\beq     \label{g33-1}
    	   g_{33} = - \frac{k^2}{m^2}\bigg[1 -\frac 1{|w|} \tanh^2 kx\bigg].
\eeq
  
  In this solution $x \in \R$, and at both extremes $x\to \pm\infty$ we have $r = \e^\beta \to 0$, i.e., 
  these are two centers of symmetry (or poles) on the $(x, \varphi)$ 2-surface, or two symmetry axes 
  from the viewpoint of 3-dimensional space. However, as follows from \rf{g33-1}, $g_{33}$ 
  is positive (hence contains CTCs) where $|\tanh kx| > |w|$, that is, at large enough $|x|$, 
  in circular regions around the two poles.
  
  By choosing another value of the integration constant $E_0$ one can make one of the poles free
  from CTCs, at the expense of enlarging the CTC region around the other pole. One of the poles 
  can even be made regular by a proper choice of the parameters. For example, choosing
  $E_0 = 2k\oo/m^2$, we obtain $E=0$ at $x = -\infty$, and it is easy to verify that the pole 
  $x = -\infty$ is then regular under the condition $k=1$.
  
\subsection*{Solution 2, \eqn{be2}}  

   For $w > 0$, $k > 0$, with \rf{be2}, for $E(x)$ we find
\beq
	E = E_0 - \frac{2 k \oo}{m^2} \coth kx.		
\eeq     
  The metric takes the form 
\bearr                                     \label{ds2}
	  ds^2 = (dt - E d\varphi)^2 - dz^2 
\nnn \cm	  
		  -  \frac {k^2} {m^2 \sinh^2 (kx)} (dx^2 + d\varphi^2), 	
\ear
  It is convenient to introduce the new coordinate $y$ by substituting
\beq
		\e^{-2kx} = 1 - \frac {2k}{y},
\eeq      
  after which we obtain 
\beq
		r^2 = \e^{2\beta} =\frac y{m^2}(y-2k). \qquad E = E_0 - \frac{2\oo}{m^2} y.
\eeq    
  The range $x >0$ is converted to $y \geq 2k$, where $y=2k$ is the axis of symmetry. The metric 
  now has the form 
\bearr                                     \label{ds2a}
	  ds^2 = (dt - E d\varphi)^2  - \frac {dy^2}{m y(y-2k)}
\nnn \inch 
		 - dz^2 - \frac{y}{m^2}(y-2k)d\varphi^2.  
\ear  
   Assuming $E_0=0$, for $g_{33}$ it is then easy to obtain the expression 
\beq       \label{g33-2}
                  g_{33} = \frac{y}{m^2} \Big(\frac yw + 2k\Big) >0, 
\eeq      
  which means that CTCs are present everywhere, and actually this space-time has an incorrect 
  signature, $(+ - - +)$ instead of $(+ - - -)$.
  
  However, with nonzero values of $E_0$ it becomes possible to get rid of CTCs in some part of space.
  Thus, choosing $E_0 $ in such a way that $E=0$ at some $y_0 > 2k$, we will obtain 
  the normal sign $g_{33} < 0$ in some range of $y$ around $y_0$.
  
\subsection*{Solution 3, \eqn{be3}}  

  In the case $w > 0$, $k = 0$, with \rf{be3}, it is convenient to use the coordinate $r = 1/(mx)$,
  and then we obtain
\bearr
	E = E_0 + \frac{2\oo}{r},
\nnn
	ds^2 = (dt - E d\varphi)^2 - \frac{dr^2}{m^2 r^2} - dz^2 - r^2 d\varphi^2,		  
\ear
  and assuming $E_0=0$, we arrive at
\beq                    \label{g33-3}
  	g_{33} = -r^2 \Big(1 - \frac {2\oo^2}{m^2}\Big) = r^2 \frac {1-w}{2w} > 0.
\eeq  	
  We again obtain a configuration with an incorrect signature, possessing CTCs at all $r$. 
  However, as in the previous case, by choosing $E_0$ so that $E=0$ at some $r=r_0$
  we can provide a CTC-free region in a thick layer around $r=r_0$.
 
\section{Wormholes}

  With the solution \rf{be4} for $r=\e^\beta$, the range of $x$ is 
  $x \in \big(\!-\!{\pi}/(2\ok), {\pi}/(2\ok)\big)$,
  where $\ok = -k >0$, and we see that $r\to \infty$ on both ends, confirming the wormhole nature of
  this configuration, where  $x=0$ is the \wh\ throat (minimum of $r$). Substituting $y = \ok \tan \ok x$, 
  we obtain the metric in the form 
\bearr                                 \label{ds4}
    	ds^2 = (dt - E d\varphi)^2 - \frac{dy^2}{m^2 (\ok^2 + y^2)} 
\nnn \inch  \cm  	
    						- dz^2 - \frac{\ok^2 + y^2}{m^2} d\varphi^2,
\ear
   where $y \in \R$, and $y=0$ is the throat; furthermore, 
\beq 					\label{E4}
		E = E_0 + \frac {2\oo}{m^2} y,
\eeq
   and for $g_{33}$ in the case $E_0=0$ (which makes the solution symmetric  with respect to $y=0$)
   it follows 
\beq                         \label{g33-4}
		g_{33} = - \frac{\ok^2}{m^2} + \frac{1-w}{2w}\,\frac{y^2}{m^2}. 
\eeq            
   The expression \rf{g33-4} shows that CTCs are absent around the throat, at $y^2 < 2\ok^2 w/(1-w)$,
   while at larger $|y|$ the CTCs emerge.
   
   Let us note that in the limit $w\to 1$, so that the fluid EoS tends to that of maximally stiff matter,
   the magnetic field disappears ($h\to 0$ according to \rf{C2}), and the whole solution tends to the
   one obtained in \cite{BBS19} for a \cy\ \wh\ with stiff matter. 
   
   As always with rotating \cy\ \wh\ solutions, these \whs\ do not have a  flat-space asymptotic behavior
   at large $|x|$, which makes it impossible to interpret them as objects that can be observed from 
   regions with small curvature. To overcome this problem, it has been suggested \cite{BLem13} to cut
   out of the obtained \wh\ solution a regular region, containing a throat, and to place it between two flat 
   regions, thus making the whole system manifestly \asflat. However, to interpret such a ``sandwich'' as 
   a single space-time, it is necessary to identify the internal and external metrics on the junction 
   surfaces $\Sigma_+$ and $\Sigma_-$, which should be common for these regions. 
   The internal region will be described in the present case by \rf{ds4}, \rf{E4}).  And since the
   internal metric contains rotation, the external Minkowski metric should also be taken in a rotating 
   reference frame.
   
  Thus we take the Minkowski metric in \cy\ coordinates, 
   $ds_{\rm M}^2 = dt^2 - dX^2 - dz^2 - X^2 d\varphi^2$, and convert it to a rotating reference frame with 
   angular velocity $\Omega = \const$ by substituting $\varphi \to \varphi + \Omega t$, so that 
\beq                                                          \label{ds_M}
	      ds_{\rm M}^2 = dt^2 - dX^2 - dz^2 - X^2 (d\varphi + \Omega dt)^2.
\eeq
  In the notations of (\ref{ds-rot}), the relevant quantities in \rf{ds_M} are 
\bearr                                                   \label{M-param}
      \e^{2\gamma} =  1 - \Omega^2 X^2,\qquad      \e^{2\beta} = \frac{X^2}{1 - \Omega^2 X^2},
\nnn
      E = \Omega X^2, \qquad      \omega = \frac{\Omega}{1 - \Omega^2 X^2}.
\ear
  This stationary metric admits matching with the internal metric at any $|X| <  1/|\Omega|$, inside the 
  ``light cylinder'' $|X| = 1/|\Omega|$ on which the linear rotational velocity coincides with the speed of 
  light.  

  Making use of the symmetry of \rf{ds4}, let us assume that the internal region is $-y_* < y < y_*$,
  so that the junction surfaces $\Sigma_{\pm}$ are situated at $y = \pm y_*$, to be identified with 
  $X_\pm = \pm X_*$ in Minkowski space, respectively, so that the external flat regions are $X < -X_*$
  and $X > X_+$. Matching  is achieved if we identify there the two metrics, so that  
\beq                                                       \label{ju-1}
		      [\beta] = 0, \quad [\mu] = 0, \quad     [\gamma] = 0, \quad [E] =0,
\eeq
  where, as usual, the brackets $[f ]$ denote a discontinuity of any function $f$ across the surface.  
  Under the conditions \rf{ju-1}, we can suppose that the coordinates $t, z, \phi$ are the same in the 
  whole space. At the same time, there is no need to adjust the choice of radial coordinates on different 
  sides of the junction surfaces since the quantities involved in all matching conditions are insensitive 
  to possible reparametrizations of $y$ or $X$.  

  Having identified the metrics, we certainly did not adjust their normal derivatives, whose jumps 
  are well known to determine the properties of matter filling a junction surface $\Sigma$ and forming
  there a thin shell. The SET $S_a^b$ of such a thin shell is calculated  using the Darmois-Israel formalism 
  \cite{darmois,israel67,BKT87}, and in the present case of a timelike surface, $S_a^b$ is related to the 
  extrinsic curvature $K_a^b$ of $\Sigma$ as
\def\tK {{\tilde K}{}}
\bearr                                                        \label{ju-2}
	S_a^b = - (8\pi G)^{-1} [\tK_a^b], \quad 	\tK_a^b := K_a^b - \delta_a^b K^c_c, 		
\ear
 where the indices $a, b, c = 0, 2, 3$.  The general expressions for nonzero components of  $\tK_a^b$
 for surfaces $x=\const$ in the metric \rf{ds-rot} are \cite{BBS19}
\bearr                              \label{tKab}
      \tK_{00} = - \e^{-\alpha+2\gamma} (\beta' + \mu'),    
\nnnv      
      \tK_{03} = -\half \e^{-\alpha} E' + E\e^{-\alpha}(\beta'+\gamma'+\mu'),
\nnnv
      \tK_{22} = \e^{-\alpha + 2\mu} (\beta'+\gamma'),
\nnnv 
      \tK_{33} = \e^{-\alpha+2\beta}(\gamma'+\mu') 
\nnn \cm      
	+ \e^{-\alpha-2\gamma}[EE'- E^2(\beta'+2\gamma'+\mu')]. 	      
\ear                   

   From \rf{tKab} it is straightforward to find $S^a_b$ on the surfaces $\Sigma_\pm$. However, our
   interest is not in finding these quantities themselves but, instead, a verification of whether or not 
   the resulting SET $S^a_b$ satisfies the WEC. Let us use for this purpose the necessary and 
   sufficient conditions obtained in a general form in \cite{BK18}, see also a detailed 
   description in \cite{ BBS19}. These conditions are                        
\bearr        \label{Wa} 
	a+c+\sqrt{(a-c)^2 +4 d^2} \geq 0,
\yyy          \label{Wb}    
	a+c+\sqrt{(a-c)^2 +4 d^2} + 2b \geq 0,
\yyy            \label{Wc}
	a + c \geq 0,          
\ear      
  where 
\bearr                 \label{abcd}
  	  a = -[\e^{-\alpha}(\beta'+\mu')],  \quad\ \,
  	  b = [\e^{-\alpha}(\beta' + \gamma')],
\nnn  	  
  	  c = [\e^{-\alpha}(\gamma'+ \mu')],\qquad\
  	  d = -[\omega]
\ear	  

  Let us discuss, for certainty, the conditions on $\Sigma_+:\ y = y_*,\ X=X_*$ with our metrics 
  \rf{ds4}, \rf{E4} and \rf{ds_M}. Among the matching conditions \rf{ju-1}, $[\mu]=0$ holds
  automatically, while to satisfy the condition $[\gamma]=0$ we will rescale the time coordinate 
  in the internal region according to
\def\tt{{\tilde t}}
\beq
		t = \sqrt{P} \tt, \qquad  P := 1 - \Omega^2 X^2
\eeq    
  and use the new coordinate $\tt$, with which, instead of $E$, we must use $E\sqrt{P}$ in all formulas. 
  The remaining two conditions \rf{ju-1} yield
\beq
	   \Omega  X^2 = \frac{2\oo}{m} y\sqrt{P}, \qquad \frac{k^2 + y^2}{m^2}  = \frac{X^2}{P},
\eeq
  where, without risk of confusion, we omit the asterisk at $X$ and $y$. With these conditions,
  there are four independent parameters of the system, for example, we can choose as such 
  parameters 
\beq
     		X,\ \  y,\ \  P,\ \  n = \frac{2\oo^2}{m} = \frac{w+1}{2w}. 
\eeq    
  The other parameters are expressed in their terms as
\bearr                        \label{param}
		\Omega=\frac{\sqrt{1{-}P}}{X}, \quad 
		\oo = \frac{ny\sqrt{P}}{X\sqrt{1-P}}, 
\nnn \cm 
		k^2 = \frac{y^2(2n - 1 + P)} {1-P}.
\ear 

    Now we can calculate the quantities \rf{abcd}, with $[f] = f_{\rm out} - f_{\rm in}$ on $\Sigma_+$:
\bearr           \label{abcd1}
		a =  \frac{y P^{3/2}-1}{PX}, \qquad
		b = \frac{1 - y\sqrt{P}}{X},
\nnn
		c = 	-\frac{1-P}{PX},\cm\
		d = \frac{ny P^{3/2}  -1 +P}{PX\sqrt{1-P}}.
\ear    
   It can be easily verified that the conditions \rf{Wa}, \rf{Wb}, \rf{Wc} are satisfied as long as
\beq  				\label{y_1}
		y \geq \frac{2-P}{P^{3/2}},
\eeq      
  in full analogy with the corresponding calculation in \cite{BBS19}. 

  We have shown that under the condition \rf{y_1} the WEC holds on $\Sigma_+$. Now, what changes 
  on the surface $\Sigma_-$ specified by $X = -X_* < 0$ and $y= -y_* <0$, where we must take 
  $[f] = f_{\rm out} - f_{\rm in}$ for any function $f$?
  As in \cite{BBS19}, it can be verified that the parameters $a,b,c$ do not change from \rf{abcd1} 
  if we replace $X$ with $|X|$ (we denote, as before, $y = y_* >0$).
  For $d = -[\omega]$ there will be another expression since, according to \rf{ju-1},
  $\Omega (\Sigma_-) = -\Omega (\Sigma_+)$, while in the internal solution 
  $\omega (\Sigma_-) = \omega (\Sigma_+)$, hence on $\Sigma_-$
\[
	d \ \mapsto \ d_- = - \frac {n|y| P^{3/2}+1 - P}{|X| P \sqrt{1-P}},
\]    
  so that $|d_-| > |d|$, making it even easier to satisfy the WEC requirements.
  As a result, the WEC holds under the same condition \rf{y_1}, providing 
  a \wh\ model which is completely phantom-free.
  
  We can also notice that from \rf{param} it follows $y_*^2 < k^2$, therefore, $y^2 < k^2$
  in the whole internal region, which is thus free from CTCs.
  
  There is one more point to bear in mind: since there is a $z$-directed magnetic magnetic field in the internal
  region, we must suppose that there are some surface currents on $\Sigma_\pm$ in the $\varphi$
  direction. Their values can be easily calculated using the Maxwell equations $\nabla_\nu F\MN = J^\mu$.
  Indeed, say, $\Sigma_+\ (x = x_*)$ separates the region where $F\MN =0$ from the one with nonzero $F\MN$,
  therefore at their junction we have $J^\mu = -\delta(x - x_*) J^\mu (x_* -0)$, so that the surface current 
  is $J^a = - J^\mu (x_*-0)\big|_{\mu=a}$. Similarly, on $\Sigma_-\ (x = -x_*)$ we obtain 
  $J^a =  J^\mu (-x_*+0)\big|_{\mu=a}$.  In our \wh\ configurations we obtain, according to 
  \rf{flu}, \rf{rho0}, \rf{ga-mu} and taking into account that $\gamma = \mu \equiv 0$,
\beq
               J^a = (J^0, 0, 0), \quad\ J^0(\Sigma_\pm) = \mp h\oo.
\eeq    
  Thus the surface currents on $\Sigma_\pm$ have only the temporal component, i.e., they are comoving 
  to the matter and current in the internal region.
    
   As is the case with the internal \wh\ solution, in the limit $w\to 1$ (hence $n\to 1$) the
   whole twice \asflat\  construction tends to the one obtained in \cite{BBS19} with a stiff matter source.
                                                                                                                              
\section{Concluding remarks}

  We have obtained a family of stationary \cyl\ solutions to the Einstein-Maxwell equations in 
  the presence of perfect fluids with $p = w\rho$, $|w| < 1$.  Some of them (Solutions 1--3) 
  contain a symmetry axis which can be made regular by properly choosing the solution parameters.
  The only geometry of closed type belongs to Solution 1, \eqs \rf{be1}, \rf{E1}, \rf{ds1}, \rf{g33-1}.
  Unlike Melvin's solution and like all other solutions with rotation, it inevitably contains a region 
  where $g_{33} > 0$, so that the coordinate circles parametrized by the angle $\varphi$ are timelike,
  violating causality. 

  The \wh\ models discussed here are of interest as new examples of phantom-free \whs\ in \GR,
  respecting the WEC. As in other known examples \cite{BK18, BBS19}, such a result is achieved 
  owing to the exotic properties of vortex gravitational fields with \cy\ symmetry, and their asymptotic
  behavior making them potentially observable from flat or weakly curved regions of space is provided
  by joining flat regions on both sides of the throat. Such a complex structure is necessary because
  asymptotic flatness at large circular radii cannot be achieved in any \cy\ solutions with rotation. 
  The present family of models with a magnetic field, parametrized by the equation-of-state parameter
  $w < 1$, tends to the one obtained in \cite{BBS19} in the limit $w\to 1$, in which the magnetic field 
  vanishes.
  
  Let us mention that other static or stationary \wh\ models with proper asymptotic behavior and 
  matter sources respecting the WEC have been obtained in extensions of \GR, such as the 
  Einstein-Cartan theory \cite{BGal15, BGal16}, Einstein-Gauss-Bonnet gravity \cite{GBo}, 
  multidimensional gravity including brane worlds \cite{BKim03, BS16}, 
  theories with nonmetricity \cite{KOL18}, Horndeski theories \cite{Su-Horn}, etc.

  One can also notice that the same trick as was used with \wh\ models, that is, joining a flat region 
  taken in a rotating reference frame, can be used as well with solutions possessing a symmetry axis. 
  It is important that in all such cases the surface to be used as a junction should not contain CTCs 
  (in other words, there should be, as usual, $g_{33} < 0$) because $g_{33} < 0$ in the admissible 
  part of flat space, while $g_{33}$ taken from the external and internal regions should be identified 
  at the junction.  In this way one can obtain completely CTC-free models of extended cosmic 
  strings with rotation. 
  
  A possible observer can certainly be located far from such an extended string or \wh\ configuration 
  and be at rest in a nonrotating frame in flat space, other than the one used for the 
  object construction.  
  
  An evident further development can be a search for other rotating configurations with electromagnetic 
  fields, possibly including radiation in different directions in the spirit of \cite{kb20-rad}, where 
  radial, azimuthal and longitudinal radiation flows were considered as sources of gravity in space-times 
  with the metric \rf{ds-rot}.  Another set of problems concerns electrostatics in the fields of extended 
  strings or \whs\ with sources including electromagnetism. As follows from \cite{sim20}, even in simpler,
  partly conical \cy\ geometries with thin shells electrostatics turns out to be rather interesting and complex.  
  
\subsection*{Acknowledgments}

  This publication was supported by the RUDN University program 5-100 and by the 
  Russian Foundation for Basic Research Project 19-02-00346. 
  The work of K.B. was also performed within the framework of the Center FRPP 
  supported by MEPhI Academic Excellence Project (contract No. 02.a03.21.0005,
  27.08.2013).  
  The work of V.K. and V.O. was supported  by the Ministry of Education 
  and Science of Russia in the framework of State Contract 9.1195.2017.6/7.

\end{document}

%% file: preamble.tex
\usepackage{amsfonts,amssymb,cite}
\usepackage{graphicx}

\def\noi{\noindent}

\newcommand{\Title}[1]{\noi {{\Large\bf #1}}\\[1ex]}

\def\Aunames#1{\noi{\bf #1}}
\def\au#1{${}^{#1}$}
\def\Addresses#1{\medskip\noi \protect
	\begin{description}\itemsep -3pt {\it #1} \end{description}}
\def\adr#1#2{\item[${}^{#1}$]{\it #2}}

\newcommand{\Abstract}[1]{\vskip 2mm \begin{center}
        \parbox{16.4cm}{\small\noi #1} \end{center}\medskip}

\def\email#1#2{\footnotetext[#1]{e-mail: #2}\addtocounter{footnote}{1}}

\def\nqq{\hspace*{-2em}}
\def\nhq{\hspace*{-0.5em}}

\def\cm{\hspace*{1cm}}
\def\inch{\hspace*{1in}}

\def\Jl#1#2{#1 {\bf #2},\ }

\def\ApJ#1 {\Jl{Astroph. J.}{#1}}
\def\CQG#1 {\Jl{Class. Quantum Grav.}{#1}}
\def\DAN#1 {\Jl{Dokl. AN SSSR}{#1}}
\def\GC#1 {\Jl{Grav. Cosmol.}{#1}}
\def\GRG#1 {\Jl{Gen. Rel. Grav.}{#1}}
\def\IJMPD#1 {\Jl{Int. J. Mod. Phys. D}{#1}}
\def\JETF#1 {\Jl{Zh. Eksp. Teor. Fiz.}{#1}}
\def\JETP#1 {\Jl{Sov. Phys. JETP}{#1}}
\def\JHEP#1 {\Jl{JHEP}{#1}}
\def\JMP#1 {\Jl{J. Math. Phys.}{#1}}
\def\NPB#1 {\Jl{Nucl. Phys. B}{#1}}
\def\NP#1 {\Jl{Nucl. Phys.}{#1}}
\def\PLA#1 {\Jl{Phys. Lett. A}{#1}}
\def\PLB#1 {\Jl{Phys. Lett. B}{#1}}
\def\PRD#1 {\Jl{Phys. Rev. D}{#1}}
\def\PRL#1 {\Jl{Phys. Rev. Lett.}{#1}}

\def\al{&\nhq}
\def\lal{&&\nqq {}}
\def\eq{Eq.\,}
\def\eqs{Eqs.\,}
\def\beq{\begin{equation}}
\def\eeq{\end{equation}}
\def\bear{\begin{eqnarray}}
\def\bearr{\begin{eqnarray} \lal}
\def\ear{\end{eqnarray}}
\def\earn{\nonumber \end{eqnarray}}

\def\nnv{\nonumber\\[5pt] {}}
\def\nnn{\nonumber\\ \lal }
\def\nnnv{\nonumber\\[5pt] \lal }

\def\yyy{\\[5pt] \lal }
\def\eql{\al =\al}

\def\dst{\displaystyle}
\def\tst{\textstyle}
\def\fracd#1#2{{\dst\frac{#1}{#2}}}
\def\fract#1#2{{\tst\frac{#1}{#2}}}
\def\Half{{\fracd{1}{2}}}
\def\half{{\fract{1}{2}}}

\def\e{{\,\rm e}}
\def\d{\partial}

\def\sign{\mathop{\rm sign}\nolimits}
\def\diag{\mathop{\rm diag}\nolimits}

\def\const{{\rm const}}
\def\eps{\varepsilon}



%% file: rot_cy-8b.bbl
\small
\begin{thebibliography}{99}   

\bibitem{exact-book}
	H. Stephani, D. Kramer,  M. A. H. MacCallum, C. Hoenselaers, and E. Herlt,
	{\it Exact solutions of Einstein's field equations,\/} 
	Cambridge Monographs on Mathematical Physics (Cambridge University Press, 2009).

\bibitem{BSW19}
	K.A. Bronnikov, N.O. Santos, and Anzhong Wang,	
	``Cylindrical systems in general relativity,'' arXiv: 1901.06561.

\bibitem{melvin}	
	M. A. Melvin, ``Pure magnetic and electric geons,''  Phys. Lett. {\bf 8}, 65-68 (1964).

\bibitem{kb79} 
	K. A. Bronnikov,  ``Static, cylindrically symmetric Einstein-Maxwell fields,''
	in {\it Problems in Gravitation Theory and Particle Theory (PGTPT)}, 
	Ed.  K. P. Staniukovich, 10th issue, p. 37-50 (Atomizdat, Moscow, 1979, in Russian). 

\bibitem{lanczos}
	C. Lanczos, 
	``\''{U}eber eine station\"{a}re Kosmologie in Sinne der Einsteinischen Gravitationstheories.'' 
	Z. Physik {\bf 21}, 73 (1924).

\bibitem{lewis}
	T. Lewis, 
	``Some special solutions of the equations of axially symmetric gravitational fields.'' 
	Proc. R. Soc.. A {\bf 136}, 176 (1932).	
	
\bibitem{vac-Lam1}
	N. O. Santos, ``Solution of the vacuum Einstein equations with nonzero
	cosmological constant for a stationary cylindrically symmetric spacetime.'' 
	Class. Quantum Grav. {\bf 10}, 2401 (1993).
	
\bibitem{vac-Lam2}	
	A. Krasinski, ``Solutions of the Einstein field equations for a rotating
	perfect fluid II: properties of the flow-stationary and and vortex-homogeneous solutions.'' 
	Acta Phys. Pol. {\bf 6}, 223 (1975).

\bibitem{vac-Lam3}	
	M. A. H. MacCallum and N. O. Santos. 
	``Stationary and static cylindrically symmetric Einstein spaces of the Lewis form.'' 
	Class. Quantum Grav. {\bf 15}, 1627 (1998).

\bibitem{BLem13}
	K. A. Bronnikov, V. G. Krechet, and Jos\'e P. S. Lemos, 
	``Rotating cylindrical wormholes.''
	\PRD {87} 084060 (2013); arXiv: 1303.2993.

\bibitem{BK15}
	K.A. Bronnikov, V.G. Krechet, 
	``Rotating cylindrical wormholes and energy conditions.''
	Int. J. Mod. Phys. A {\bf 31}, 1641022 (2016); arXiv: 1509.04665.

\bibitem{erices}
	C. Erices and C. Martinez, 
	``Stationary cylindrically symmetric spacetimes with a massless scalar field
	and a nonpositive cosmological constant.''
	Phys. Rev. D {\bf 92}, 044051 (2015).

\bibitem{van_stockum}
	W. J. van Stockum, 
	``The gravitational field of a distribution of particles rotating about an axis of symmetry.'' 
	Proc. Roy. Soc. Edinburgh  A {\bf 57}, 135 (1937).

\bibitem{bonnor09}
	W. B. Bonnor and B. R. Steadman, 
	``A vacuum exterior to Maitra's cylindrical dust solution.'' Gen. Rel. Grav. {\bf 41}, 1381 (2009).

\bibitem{iva02c}
	B. V. Ivanov, ``The general double-dust solution,'' gr-qc/0209032

\bibitem{iva02b}
	B.V. Ivanov,	
	``Rigidly rotating cylinders of charged dust.''
	\CQG {19} 5131 (2002); gr-qc/0207013.

\bibitem{santos82}
	N. O. Santos and R. P. Mondaini,
	``Rigidly rotating relativistic generalized dust cylinder.'' 
	Nuovo Cim. B. {\bf 72}, 13 (1982)

\bibitem{hoen79}
	C. Hoenselaers and C. V. Vishveshwara, A relativistically rotating fluid cylinder.
	\GRG {10} 43-=51 (1979).

\bibitem{davids97}
	W. Davidson, ``Barotropic perfect fluid in steady cylindrically symmetric rotation.''
	\CQG {14} 119 (1997).

\bibitem{davids00}
	W. Davidson, ``A cylindrically symmetric stationary solution of Einstein's
	equations describing a perfect fluid of finite radius.'' 
	\CQG {17} 2499 (2000). 

\bibitem{skla99}
	D. Sklavenites, 
	``Stationary perfect fluid cylinders.''
	\CQG {16} 2753 (1999).

\bibitem{iva02a}		
	B.V. Ivanov,
	``On rigidly rotating perfect fluid cylinders.''
	\CQG {19} 3851 (2002); gr-qc/0205023.

\bibitem{anis1}
	P. S. Letelier and E. Verdaguer.
	``Anisotropic fluid with SU(2) type structure in general relativity: A model of localized matter.''
	\JMP {28} 2431 (1987). 

\bibitem{anis2}
	L. Herrera, G. Le Denmat, G. Marcilhacy, and N. O. Santos.
	``Static cylindrical symmetry and conformal flatness.''
	Int. J. Mod. Phys. D 14 (2005) 657; gr-qc/0411107.

\bibitem{anis3}
	F. Debbasch, L. Herrera, P.R.C.T. Pereira, N.O. Santos,
	``Stationary cylindrical anisotropic fluid.''
	Gen. Rel. Grav. {\bf 38}, 1825   (2006); gr-qc/0609068.
	
\bibitem{anis4}	
	S. V. Bolokhov, K. A. Bronnikov, and M. V. Skvortsova, 
	``Rotating cylinders with anisotropic fluids in general relativity,''
	\GC {25} 122 (2019); arXiv: 1904.06727.

\bibitem{BK18}
	K. A. Bronnikov and, V. G. Krechet, 	
	``Potentially observable cylindrical wormholes without exotic matter in GR.''
	Phys. Rev. D {\bf 99}, 084051 (2019); arXiv: 1807.03641.

\bibitem{BBS19}
	K. A. Bronnikov, S. V. Bolokhov, and M. V. Skvortsova, 
	``Cylindrical wormholes: a search for viable phantom-free models in GR.'' 
	Int. J. Mod. Phys. D {\bf 28}, 1941008 (2019); arXiv: 1903.09862.

\bibitem{kr2}
	V. G. Krechet, 
	Topological and physical effects of rotation and spin in the general relativistic theory of gravitation,
	Izvestiya Vuzov, Fiz. {\bf No 10}, 57 (2007);
	Russ. Phys. J. {\bf 50}, 1021 (2007).

\bibitem{kr4}
	V. G. Krechet and D. V. Sadovnikov, 
	Spin-spin interaction in general relativity and induced geometries with nontrivial topology.
	\GC {15} 337 (2009); arXiv: 0912.2181.

\bibitem{darmois} 
	G. Darmois, Les \'equations de la gravitation einsteinienne, 
	in: {\it M\'emorial des Sciences Mathematiques} 
	(Gauthier-Villars,  Paris, 1927), vol. 25.

\bibitem{israel67}
	W. Israel, Singular hypersurfaces and thin shells in general relativity.
	Nuovo Cim. B {\bf 48}, 463 (1967). 

\bibitem{BKT87}
	V. A. Berezin, V. A. Kuzmin, and I. I. Tkachev, 
	Dynamics of bubbles in general relativity.
	\PRD {36} 2919 (1987). 

\bibitem{ws-book}
	K. A. Bronnikov and S. G. Rubin, {\it Black Holes, Cosmology and Extra Dimensions}
	(World Scientific, 2013).

\bibitem{BGal15}
	K.A. Bronnikov, A.M. Galiakhmetov,
	Wormholes without exotic matter in Einstein-Cartan theory,
	\GC {\bf 21} 283 (2015); arXiv: 1508.01114.

\bibitem{BGal16}
	K.A. Bronnikov, A.M. Galiakhmetov,
	Wormholes without exotic matter in Einstein-Cartan theory,
	\PRD {\bf 94} 124006 (2016); arXiv: 1607.07791.

\bibitem{GBo}
	H. Maeda, M. Nozawa,
        Static and symmetric wormholes respecting energy conditions in Einstein-Gauss-Bonnet gravity.
	\PRD {78} 024005 (2008).

\bibitem{BKim03}
	K.A. Bronnikov,  S.-W. Kim,
         Possible wormholes in a brane world.
	\PRD {67} 064027 (2003);	gr-qc/0212112.

\bibitem{BS16}
	K.A. Bronnikov, M.V. Skvortsova,
	Wormholes leading to extra dimensions
	Grav. Cosmol. {\bf 22}, 316 (2016). 			

\bibitem{Su-Horn}
	Sergey V. Sushkov, R V. Korolev, 			
	Scalar wormholes with nonminimal derivative coupling,
	Class. Quantum Grav. {\bf 29}, 085008 (2012); arXiv: 1111.3415.

\bibitem{KOL18}
	V. G. Krechet, V. B. Oshurko, and M. N. Lodi,
	Induced nonlinearities of the scalar field and wormholes in the metric-affine theory of gravity,
	\GC {24} 186--190 (2018). 

\bibitem{kb20-rad}
	K. A. Bronnikov, 
	String clouds and radiation flows as sources of gravity in static or rotating cylinders,
	Int. J. Mod. Phys. A {\bf 35}, 2040004 (2020); arXiv: 1909.09409.
	
\bibitem{sim20}	
	E. Rub\'{\i}n de Celis and C. Simeone,
	Electrostatics and self-force in asymptotically flat cylindrical wormholes,
	Eur. Phys. J. C {\bf 80}, 501 (2020); arXiv: 2004.08506.	  
	
\end{thebibliography}
